\documentclass[%
 reprint,
superscriptaddress,
 amsmath,amssymb,
 aps,
]{revtex4-2}

\usepackage{graphicx}
\usepackage{dcolumn}
\usepackage{bm}

\usepackage[mathlines]{lineno}

\begin{document}


\title{Known by the company we keep:
`Triadic influence' as a proxy for compatibility in social relationships}

\author{Miguel Ruiz-García}
\affiliation{Departamento de
Matem\'aticas, Universidad Carlos III de Madrid, 28911 Legan\'es, Madrid,
Spain}
\author{Juan Ozaita}%
\affiliation{Departamento de
Matem\'aticas, Universidad Carlos III de Madrid, 28911 Legan\'es, Madrid,
Spain}
\affiliation{Grupo Interdisciplinar de Sistemas Complejos (GISC),  Universidad Carlos III de Madrid, 28911 Legan\'es, Madrid,
Spain}
\author{Mar\'\i a Pereda}
\affiliation{Grupo Interdisciplinar de Sistemas Complejos (GISC),  Universidad Carlos III de Madrid, 28911 Legan\'es, Madrid,
Spain}
\affiliation{Grupo de Investigación Ingeniería de Organización y Logística (IOL), Departamento Ingeniería de Organización, Administración de empresas y Estadística, Escuela Técnica Superior de Ingenieros Industriales, Universidad Politécnica de Madrid, 28006 Madrid, Spain}
\author{Antonio Alfonso}
\affiliation{LoyolaBehLAB \& Department of Economics, Universidad Loyola Andalucía, 14004 Córdoba, Spain}
\author{\mbox{Pablo Bra\~nas-Garza}}
\affiliation{LoyolaBehLAB \& Department of Economics, Universidad Loyola Andalucía, 14004 Córdoba, Spain}
\author{Jos\'e A. Cuesta}
\affiliation{Departamento de
Matem\'aticas, Universidad Carlos III de Madrid, 28911 Legan\'es, Madrid,
Spain}
\affiliation{Grupo Interdisciplinar de Sistemas Complejos (GISC),  Universidad Carlos III de Madrid, 28911 Legan\'es, Madrid,
Spain}
\affiliation{Instituto de Biocomputaci\'on y F\'\i sica de Sistemas Complejos (BIFI), Universidad de Zaragoza, 50018 Zaragoza, Spain}
\author{Angel S\'anchez}
\affiliation{Departamento de
Matem\'aticas, Universidad Carlos III de Madrid, 28911 Legan\'es, Madrid,
Spain}
\affiliation{Grupo Interdisciplinar de Sistemas Complejos (GISC),  Universidad Carlos III de Madrid, 28911 Legan\'es, Madrid,
Spain}
\affiliation{Instituto de Biocomputaci\'on y F\'\i sica de Sistemas Complejos (BIFI), Universidad de Zaragoza, 50018 Zaragoza, Spain}

\date{\today}

\begin{abstract}
Networks of social interactions are the substrate upon which civilizations are built. Often, we create new bonds with people that we like or feel that our relationships are damaged through the intervention of third parties. Despite their importance and the huge impact that these processes have in our lives, quantitative scientific understanding of them is still in its infancy, mainly due to the difficulty of collecting large datasets of social networks including individual attributes. In this work, we present a thorough study of real social networks of 13 schools, with more than 3,000 students and 60,000 declared positive and negative relations, including tests for personal traits of all the students. We introduce a metric---the `triadic influence'---that measures the influence of nearest-neighbors in the relationships of their contacts. We use neural networks to predict the relationships and to extract the probability that two students are friends or enemies depending on their personal attributes or the triadic influence. We alternatively use a high-dimensional embedding of the network structure to also predict the relationships. Remarkably, the triadic influence (a simple one-dimensional metric) achieves the highest accuracy at predicting the relationship between two students. We postulate that the probabilities extracted from the neural networks---functions of the triadic influence and the personalities of the students---control the evolution of real social networks, opening a new avenue for the quantitative study of these systems.
\end{abstract}

\maketitle


Positive relationships help individuals thrive in society, whereas negative ones can jeopardize our chances of success and happiness. Social relationships arise from interactions between individuals and have been studied on different time scales and contexts \cite{jackson2010,easley2010}. As a result, social networks are formed, with individuals as nodes and interactions as links \cite{wasserman1994}, and they can be studied and characterized using a complex network approach \cite{newman2010} in order to assess the many implications of social structure in our lives \cite{dunbar2020}.  A great deal of research has been carried out on social networks by aggregating the interactions that occur over a certain period of time to define links, starting from the pioneering work of Moreno \cite{moreno1934}.  However, such an approach does not capture the dynamics of relationships, which is necessary to advance our understanding of the field \cite{granovetter1973}. Large efforts have been devoted to this question in recent years, mainly using empirical data with different degrees of time resolution, such as, e.g., letter exchanges \cite{oliveira2005}, mobile phone communications \cite{onnela2007,urena2020}, spatial mobility \cite{brockmann2006}, or face-to-face interactions \cite{cattuto2010,leecaster2016,gelardi2020} (see also Ref.\ \cite{holme2012} for a review). All these analyses have led to many interesting insights on the evolution of relationships, but the issue of the mechanisms that explain how/why these relationships are created and evolve remains elusive. 

Several models have been proposed to explain different aspects of the empirical observations. The first attempts were devoted to reproduce some of the structural properties observed in social networks, such as the small world phenomena \cite{watts1998} or the rich-get-richer effect \cite{barabasi1999, bianconi2001}. Starnini {\em et al.} \cite{starnini2013} proposed a simple model based on random walks and individual attractiveness to describe face-to-face interactions. For social networks, Jin {\em et al.}  \cite{jin2001} studied networks with exponential decay of tie strengths to represent friendships. Other approaches have resorted to exponential random graph models \cite{hanneke2010} or stochastic actor-oriented models \cite{snijders2010}. Finally, regression models that incorporate a selection of individual traits have also been considered for online social networks \cite{peter2005}. Still, none of these approaches sheds light on friendship formation in real life, taking into account the characteristics of the individuals and how some relationships can influence others.

In this paper, we contribute towards the understanding of friendship formation by adopting a different point of view, namely that of link prediction in networks 
\cite{liben-nowell2007}. The problem of link prediction, as originally formulated, is about temporal networks: given the graph of connections between certain entities or nodes during some interval, the task is to predict the set of links in a later interval. Notwithstanding this definition, the same idea applies to many different situations, such as recommendation systems \cite{lu2012}, bioinformatics \cite{airoldi2008}, scientific collaboration networks \cite{newman2001}, criminal networks \cite{berlusconi2016}, or even estimating the reliability of network data \cite{guimera2009}, to name a few. In the case of online social networks, link prediction has been considered, for example, by Song {\em et al.} \cite{song2009} or Hao \cite{hao2019} (see Ref.~\cite{kumar2020} for a review). Much less has been explored regarding real-world social networks, in particular friendship networks \cite{tamarit2019}, due to the difficulty of collecting data on reasonably complete social networks that include personal attributes in real settings. For this reason, the discussion has been devoted in many cases to ego-networks (i.e., data on disconnected individuals who mentioned their friends) and to the meaning of friendship \cite{buijs2022}. 

In this work we study social networks collected in $13$ complete high schools in Spain, containing more than $3,000$ individuals and $60,000$ declared relationships between them.  All students completed tests including information about their self-declared gender, cognitive results, and other variables that measured their \textit{selfishness/prosociality}. Performing link prediction on this data, we are able to extract the probability that two students will be friends/enemies depending on their personalities. We also studied how this probability is affected by other relationships, defining a metric that we have termed \emph{triadic influence}. Although we analyze static networks, our results suggest that the probabilities that we extract determine the mechanisms that control the initial formation of relationships and the evolution of the whole social network.

\section*{Results}

\begin{figure}
\centering
\includegraphics[width=\linewidth]{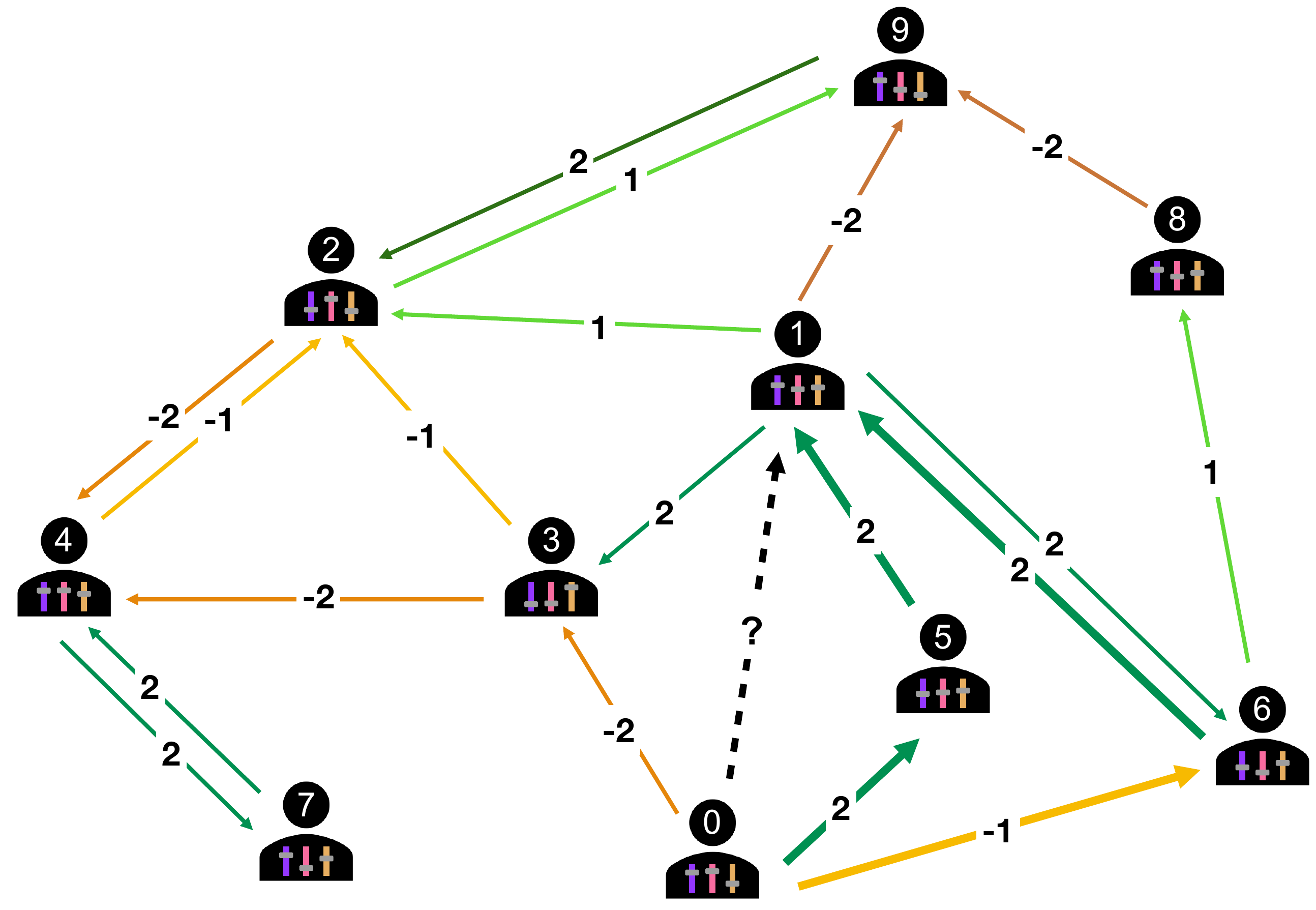}
\caption{Diagram of a social network that includes personality traits and computation of the triadic influence. To predict the relationship from node $0$ to node $1$ we can use the individual features of both students (represented by the sliders within their body) and / or the triadic influence $I_{01}$. The directions of these relationships are marked by arrows going from the nominator to the nominee, whereas the weight/intensity is represented with colors and edge labels (dark green: close friend, green: friend, yellow: dislike, orange: enemy). The thick arrows highlight the relationships that enter the calculation of $I_{01}$. To compute $I_{01}$ we select all directed paths of length $2$ from node $0$ to node $1$ ($0\rightarrow\text{node}\rightarrow 1$). In this example, they are 0-5-1 and 0-6-1. The path 0-3-1 is not a directed path (the direction of the edges is $0\rightarrow 3\leftarrow 1$) and therefore is not included in the calculation of $I_{01}$. Thus, $I_{01}=w_{05}w_{51}+w_{06}w_{61}=2\cdot 2+(-1)\cdot 2=2$.}
\label{fig_diagram}
\end{figure}

Data collection was carried out in $13$ schools in different areas of Spain, with a total of $3,395$ students. They were asked to choose with whom they were related within their school by picking names from a school list. Then they had to rate the relationship as very bad, bad, good, or very good, which we codified as $-2$, $-1$, $+1$, and $+2$, respectively. We recovered $60,566$ declared relationship, see Supporting Information (SI) for more details. In addition, we also collected data on the students' gender (self-reported), cognitive skills (measured by the cognitive reflection test, CRT), and their prosociality (see Methods for details on these individual features). With this information, we build a directed weighted network, with each link representing a relationship that goes from the nominator to the nominee---two nodes can be connected by links in both directions---weighted by the reported rating. Additionally, each node represents one student and has his/her individual attributes (gender, CRT and prosociality). Figure~\ref{fig_diagram} presents a sketch of the kind of social network that we will study. We have included several figures studying the structure of the social networks in the Supporting Information, see figures S1 to S3.

In this work, we study the correlations between the personal features of both students and the type of relationship between them, as well as the influence of other students on that relationship. We have used artificial neural networks to perform link prediction within our dataset from two complementary viewpoints: the first one focuses on the local structure, using the personality traits of both students and the influence of the nearest-neighbors as described in the next section; the second one uses only the structural information of the network -- the undirected and unweighted graph -- to predict relationships. In what follows, we discuss these two approaches separately.

\subsection*{Predicting with the personality traits and the influence of the nearest-neighbors}

Figure \ref{fig_diagram} shows a sketch of the social network with all the information available to perform link prediction. It shows the students (nodes) with their traits (sliders) and relationships of different types between them. In this section, we use only local properties of the network to predict the relationship between two students, namely the individual features of both students (e.g. nodes $0$ and $1$ in Fig. \ref{fig_diagram}) and the directed weighted paths of length $2$ between them. Specifically, we define a variable that we term \textit{triadic influence} as $I_{ij}\equiv\left(W^2\right)_{ij}=\sum_k w_{ik}w_{kj}$, where $w_{ik}$ is the weight of the link that goes from node $i$ to node $k$ (see Fig.~\ref{fig_diagram} for an example). The triadic influence condenses into one scalar the influence of third parties; e.g., if node $i$ declares node $k$ as a friend and $k$ does the same with $j$, it adds a positive number to $I_{ij}$ (your friend's friends are likely to be your friends), whereas a path containing links of opposite sign will lead to a negative contribution (your enemy's friends or your friend's enemies are likely to be your enemies). $I_{ij}$ adds up the contribution from all directed paths of length $2$ between $i$ and $j$.

For simplicity, we will train a neural network to correctly classify all relationships in the network into two classes: friends and enemies (see Methods for more details). We used different combinations of the triadic influence and the individual characteristics of the students as input for the deep neural network (NN) and trained it to output the correct value for each relationship in the training dataset (see Methods for a full description of the neural network and the training process). With our procedure, we obtain the probability that two students relate through a relationship belonging to one of the two classes (friends or enemies) as a function of the corresponding inputs. To avoid using a misleading metric of performance, since our classes are unbalanced---there are many more declared friends than enemies---we assess the performance of our method using the balanced accuracy on the test dataset. To compute it, after training the NN we feed it with all relations in the test dataset and assign the label ``friend'' or ``enemy'' to the class with the highest probability. The balanced accuracy is then computed as
\begin{equation*}
    \textrm{bAcc} = \frac{1}{2} \left( \frac{N_+^C}{N_+^T}+\frac{N_-^C}{N_-^T} \right),
\end{equation*}
where $N_\alpha^C$ is the number of samples belonging to class $\alpha$ ($+$ friend, or $-$ enemy) that were correctly classified from the total number of samples belonging to that class ($N_\alpha^T$). This is more informative than other performance metrics because if either the NN classified everything in the same class or guessed at random, we would obtain $\textrm{bAcc}=1/2$ regardless of the number of samples in each class, whereas if all relations were correctly predicted, then $\textrm{bAcc}=1$ (see Methods).

\begin{figure}
    \centering
    \includegraphics[width=\linewidth]{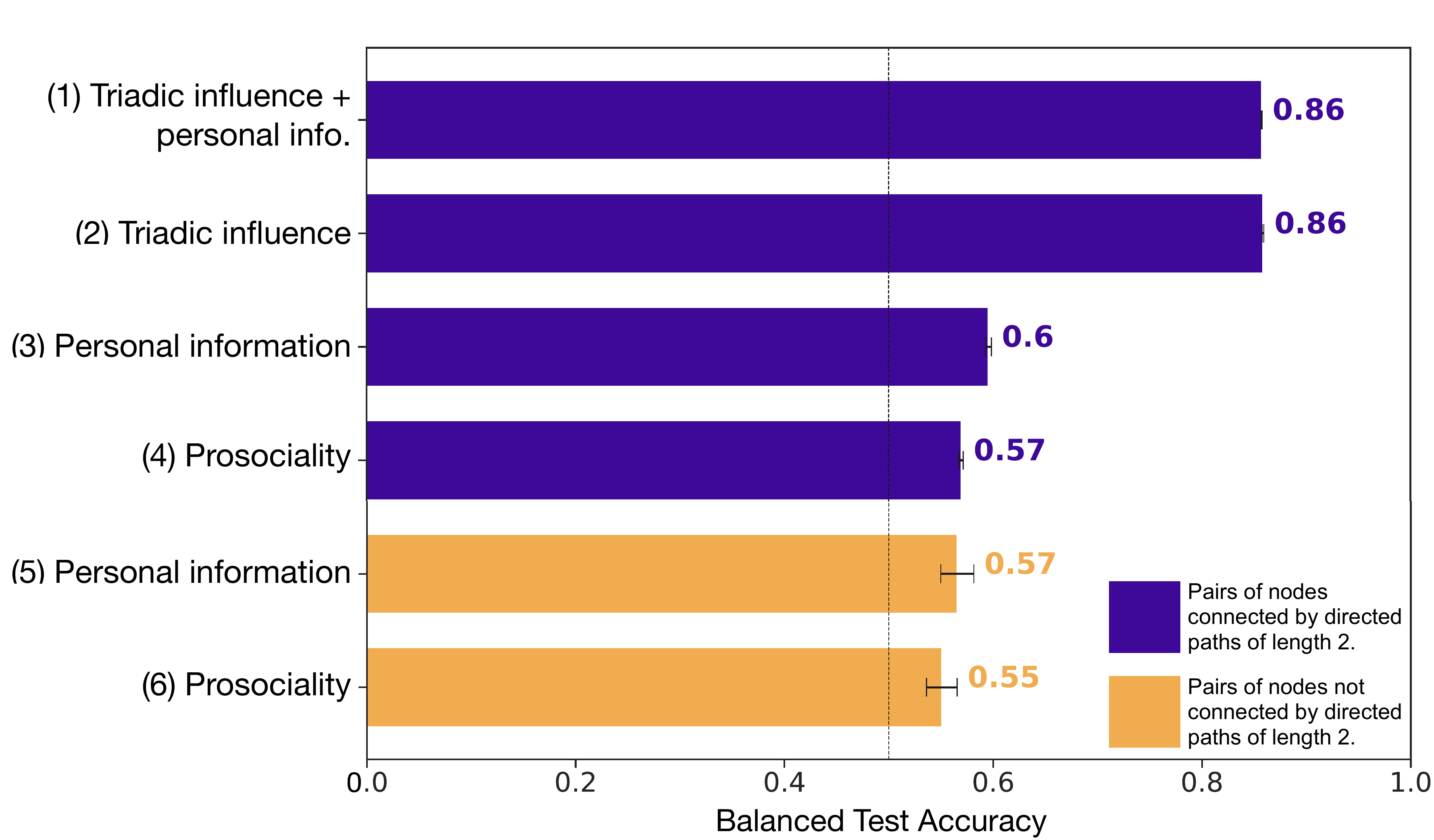}
    \caption{Balanced test accuracy for different choices of information used to train the NN. Purple bars correspond to relationships where there is at least one directed path of length $2$ from $i$ to $j$ ($(A^2)_{ij}>0$, $A_{ij}$ being the adjacency matrix of the network). We train the classifier using four sets of predictors: (1) triadic influence and personal information (gender, CRT, and prosociality), (2) triadic influence alone, (3) personal information alone, (4) just students' prosociality. In all four cases we trained $10$ different NN with random initializations and show here the mean $\textrm{bAcc}$. Yellow bars correspond to the $\textrm{bAcc}$ for relationships that have no directed paths of length $2$. In this case, we use just two sets of predictors: (5) personal information and (6) students' prosociality. These cases use $10$-fold cross-validation to estimate the performance of the prediction. Error bars represent the standard error of the mean in all cases. 
    }
    \label{fig_acc}
\end{figure}

Figure~\ref{fig_acc} collects the accuracies achieved using the NN to predict the relationships between students with different combinations of predictors. We first study relationships $i\to j$ with non-zero triadic influence (i.e., with at least one directed path of length $2$ from $i$ to $j$; see Fig.~\ref{fig_diagram} and Methods for more details). The results are shown in the four upper bars of Fig.~\ref{fig_acc} (see the SI for the distribution of relationships per number of directed paths of length $2$, Fig.~S2). We train the classifier using four sets of predictors: (1) triadic influence and personal information (gender, CRT, and prosociality) of the pair of nodes, (2) triadic influence, (3) personal information, and (4) only students' prosociality. Just as a clarification, in case (1) we use as input for the NN the triadic influence (a scalar) and the individual traits of both students (a $6$-dimensional array) to predict the correct label of that relation (friend or enemy). See Methods for a detailed explanation about how the value of the considered features: gender, CRT, and prosociality, are gathered and computed.

The highest balanced accuracy, $86$\%, is achieved using the triadic influence as input, either in combination with personal information of both students (1) or alone (2). It is remarkable that such a high accuracy for the prediction of the nature of a relationship (friend/enemy) can be obtained with just a scalar (the triadic influence), and that a 6-dimensional array containing information about both students' characteristics does not improve on that. This suggests that the information on the individual features is already encoded in the network of relationships and captured by the triadic influence. The rationale for this is the following: if individuals $i$ and $j$ are both very prosocial, then $i$ will be friends with $j$, as well as with many other intermediate individuals, with high probability; and these intermediate individuals will very likely be friends with $j$, since $j$ is also very prosocial. Hence, $I_{ij}$ already encodes information about the prosociality of $i$ and $j$. This suggests that $I_{ij}$ may act as a proxy for personal compatibility when individual traits are not available.

On the other hand, using only the personal traits of both students (3) yields $\textrm{bAcc}=60$\%. We studied the three attributes (gender, CRT, and prosociality) separately, and prosociality turned out to be the most predictive. In fact, using only students' prosociality to predict their relationship (4) already yields $\textrm{bAcc}=57$\%. Both figures are above the accuracy that a random guess would reach ($50$\%). Note that prosociality is calculated with students' answers to three simple questions (see Methods). It is really remarkable that such a simple metric is already predictive for the nature of the social relationship between two individuals.

Finally, we study separately the relationships that do not have directed paths of length $2$ connecting $i$ to $j$ (i.e.~$\left(A^2\right)_{ij}=0$, with $A_{ij}$ the adjacency matrix of the network); therefore, there is no triadic influence between $i$ and $j$. These results are shown by the two bottom bars of Fig.~\ref{fig_acc}. Since this dataset is much smaller ($2$\% of all relationships, i.e.~$1,211$ out of a total of $60,566$; see the SI Fig.~S2 for more details), we assess the performance of the classifier using 10-fold cross-validation to ensure that our results are robust. We study two sets of predictors: (5) the complete personal information of the students (gender, CRT and prosociality) and (6) just the prosociality. The mean $\textrm{bAcc}$ for the 10 realizations within 10-fold cross-validation is $57$\% for (5) and $55$\% for (6). Note that the mean $\textrm{bAcc}$ seems to decrease compared to the case when $\left(A^2\right)_{ij}>0$ (purple bars), although the significance of this difference is low given that error bars corresponding to cases (3) and (5), and (4) and (6) either overlap or are very close.

\subsection*{Interpreting the probabilities learned by the neural network}

\begin{figure}
    \centering
    \includegraphics[width=\linewidth]{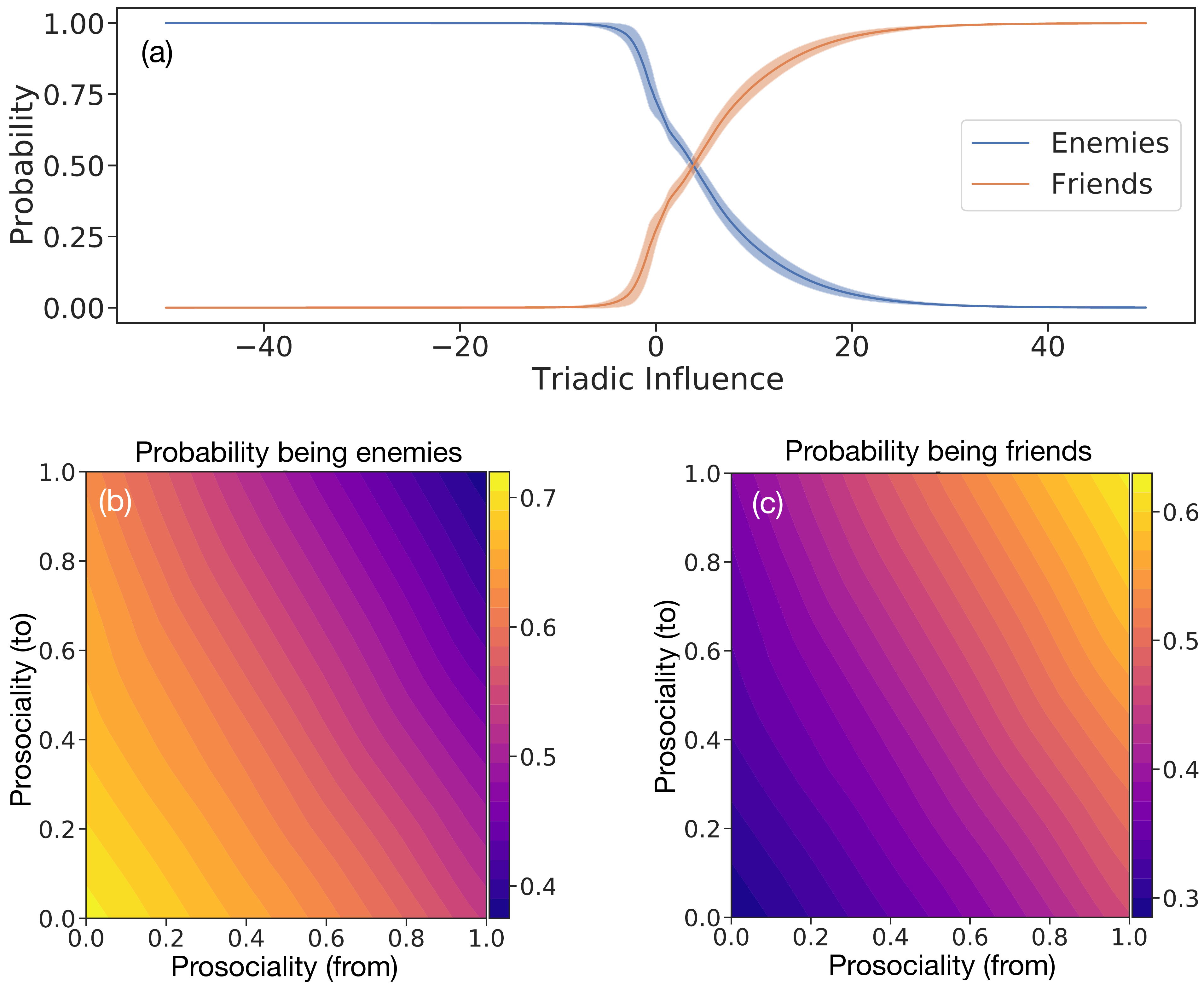}
    \caption{Probabilities of being friends/enemies as a function of the triadic influence and prosociality. Panel (a) shows the probability learned by the NN as a function of the triadic influence. We performed 10 simulations that led to the accuracy shown in the (2) bar in Fig. \ref{fig_acc}. Continuous lines in panel (a) correspond to the mean, whereas the shaded area correspond to one standard error of the mean. Panel (b) and (c) display the mean probabilities learnt by the 10 NN used in Fig. \ref{fig_acc} (4), they show the probability of having a friendly/enmity relationship as a function of the prosociality of both students, the nominator (from) and nominee (to). Both probabilities are normalized to $1$. 
    }
    \label{fig_prob_three_panels}
\end{figure}

It is important to note that until now we have chosen to assess the performance of our prediction using $\textrm{bAcc}$ for the sake of simplicity. However, the NN learns more than this; in particular, it learns to predict the probability that a relationship belongs to each of the classes in the dataset (see Methods for a detailed explanation on how this is achieved through the minimization of the cross-entropy loss function). The great advantage of using low-dimensional inputs is that we can interpret what the NN is learning. We can plot the probability that a sample belongs to a class (friend/enemy) as a function of the different predictors. In Fig.~\ref{fig_prob_three_panels} (a) we plot this probability as a function of the triadic influence. We use the $10$ different NNs trained for Fig.~\ref{fig_acc} (2) and plot the average probability of being friends and enemies for a pair  of students with a given triadic influence. The colored area around both curves represents the standard deviation of the probabilities. The probability of being friends saturates to $1$ when the triadic influence $I_{ij}\gg 1$, and drops to $0$ if the triadic influence $I_{ij}\lesssim 0$ (the probability of being enemies is the complementary because both add up to $1$). The probability curves of being friends and being enemies cross around $I_{ij}\approx 5$. Note that this is the only information used when computing the accuracy $\textrm{bAcc}$, because we identify each relationship with the most probable one, as predicted by the NN. However, the probabilities learned by the neural network (which minimize the cross-entropy loss, see Methods) contain much more information and could be used to generate ensembles of social networks or to simulate their evolution using stochastic Markov chains. It is worth mentioning that, although the probability curves change abruptly around $I_{ij}\approx 0$, this change slows down as the triadic influence increases, thus displaying an asymmetric behavior on both sides of the crossing point $I_{ij}\approx 5$.

Figures~\ref{fig_prob_three_panels} (b) and (c) display the probability of being enemies and friends, respectively, as a function of the prosociality of both students (nominator/nominee), averaged over the $10$ simulations used for case (4) of Fig.~\ref{fig_acc}. Similarly to the case of the triadic influence, even though $\textrm{bAcc}$ is fully determined by the curve where the probability is $0.5$, the profiles shown in these figures convey much more information. In particular, we can see that the probability that two students with $0$ prosociality are enemies is $70\%$, which is in line with what one would expect: selfish people declare to have more enemies and are declared enemies more often than altruists (see the Supporting Information, where this can also be directly observed in the raw data, Fig.~S4). Alternatively, two highly prosocial students are friends with a probability higher than $60\%$. Note also that both colormaps are approximately symmetric with respect to the diagonal. This implies reciprocity: the probability that $i$ declares $j$ as a friend is approximately the same as the probability that $j$ does the same with $i$.

\subsection*{Predicting with the structural information of the social network alone}

\begin{figure}
\centering
\includegraphics[width=\linewidth]{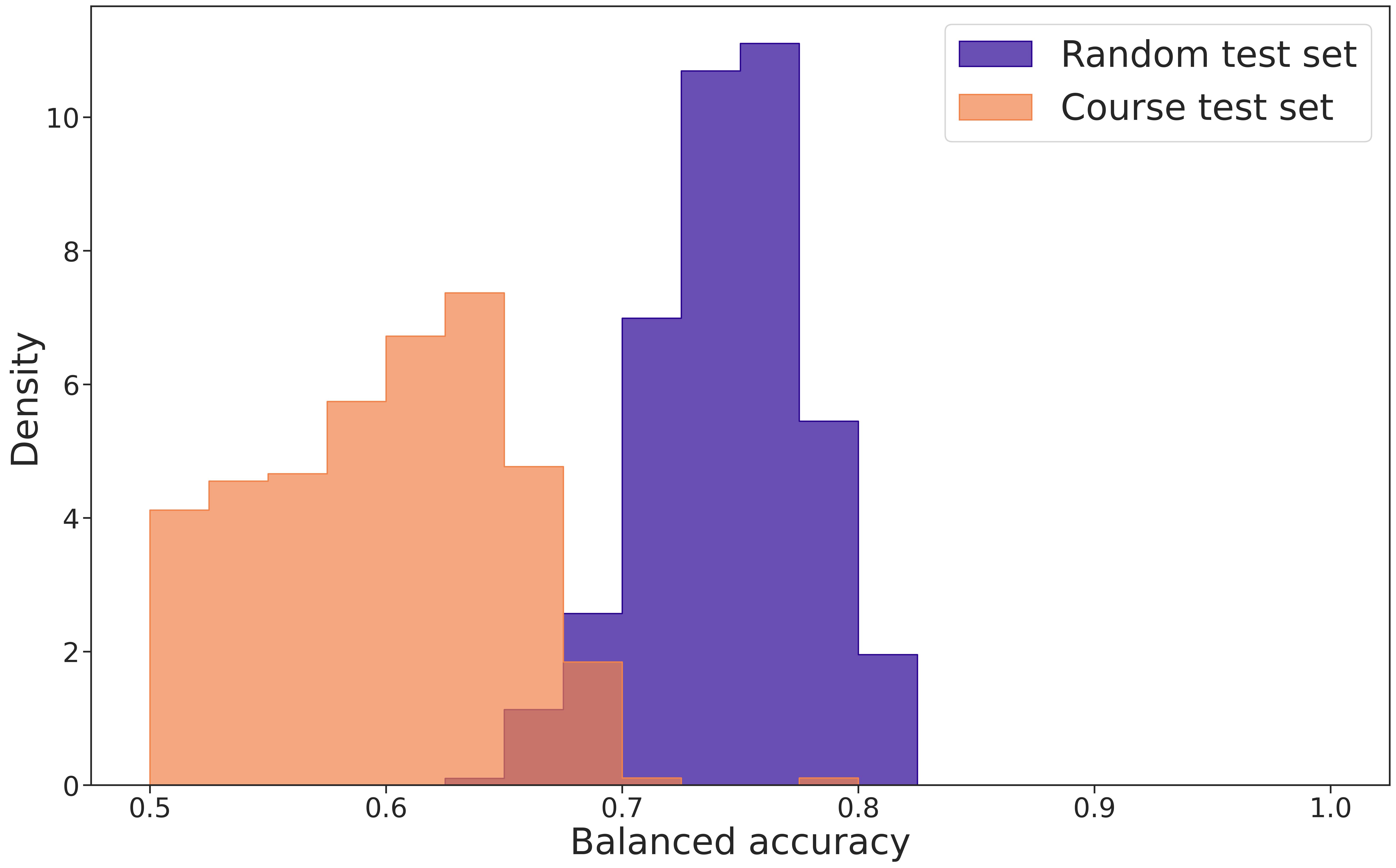}
\caption{\label{fig:AUC}Distribution of balanced accuracy for the 13 high schools. Each histogram is composed of a sample of $N=390$ points, which are different simulations for the same treatment. The histograms are normalized so that the area under the curve is $1$. Purple (dark) histogram represents \emph{treatment I} (see Methods), where we use a random pick of nodes as test set. Orange (light) histogram represents \emph{treatment II}, where we pick a specific course from the high school as the test set. The darker orange in the figure represents the overlapping in the distribution. The same figure for the Random Forest method can be checked in Fig.\ref{fig:AUC_RF}.}
\label{fig_global}
\end{figure}

In the previous sections we use local information---individual features and triadic influence---to predict relationships. Complementary to this, in this section we will attempt to make the same predictions using only the structure of the network---excluding weights, link directions, and individual features---hoping to shed light on the role played by the structure of the network for the creation of different relationships. We will merge labels $\{+1,+2\}$ into a unique ``friends'' label, and labels $\{-1,-2\}$ into a unique ``enemies'' label, so that predictions can be binary. In order to do that we will create \emph{node embeddings} by assigning to each node a $d$-dimensional array of features---which will replace the array of individual features used in the previous section. $128$-dimensional embeddings are created with Node2Vec \cite{grover2016node2vec}, an algorithm that explores the neighborhood of each node using biased random walks (see Methods for more details and figures~S5 and S6 of the SI). The embeddings of all nodes are then used as inputs to train different models, in order to predict the relationships in the network. We show here the case where we train a neural network, although we have also used Random Forests (see the SI, Fig.~S7) obtaining similar results.

We create the embeddings for all nodes once and keep them throughout. We then train a neural network to predict the relationship between pairs of students (friends/enemies) using both their embeddings as input. This is akin to using the individual features of both students in the previous section, only this time embeddings encode information about the environment surrounding each node. We have trained and tested the neural network using two alternative treatments: in \emph{treatment I} we have chosen at random $20\%$ of the relationships from all high schools as the test dataset, and trained the neural network using the rest of the relationships; in \emph{treatment II} we have created a test dataset with all the relationships inside one specific age level from one high school, and trained the model using all the other relationships. For treatment I, we trained the neural network 390 times, every time changing the train and test datasets as well as the initialization of the neural network (the embeddings do not change). For treatment II, there are 39 different age levels within the 13 high schools that we study, and we trained 10 different neural networks for each course---390 simulations in total. 

The results corresponding to treatments I and II are summarized in Fig. \ref{fig_global}. In this figure we show the accuracy as a histogram after carrying out treatments I and II for the 390 simulations---blue and orange bars, respectively. For treatment I, where we train and test on random relationships, the average accuracy is $\sim 75\%$, and the accuracy is always above $60\%$ (purple bars). However, when we test on a complete age level that was excluded from the training dataset the performance degrades (treatment II), the mean accuracy is now $\sim 60\%$ (orange bars), and there are many instances where the model is not doing better than a dummy model (bAcc $\sim 50\%$).

The fact that the model has predictive power using only structural information shows that there is a structural difference between the environments of friendly or adversarial relationships. Besides, since the predictive power of the model decreases when testing on an isolated age level, this suggests that the structure of most age levels contain specific information that is not present in the rest of the data.  In other words, some of the information that is contained within an age level, and stored in the embedding of the students, seems to be exclusive to that level.

\section*{Discussion}


In this paper, we have applied techniques for link prediction to gain insight into the mechanisms behind the formation and evolution of social networks. This has been possible due to the large amount of data that we have collected, comprising individual features of more than $3,000$ students as well as their corresponding network of personal relationships---over $60,000$ connections. The picture of the network dynamics that emerges from our work is as follows. Some initial relationships appear between pairs of students, promoted by their prosocial stance. As a matter of fact, we have shown that the prosociality of both students by themselves are capable of predicting isolated relationships significantly better than a pure random guess. This is actually a very strong claim, because many of those initial relationships are now hidden among many other relationships that emerged afterwards, and the isolated ones that we can find now are probably very sensitive to noise or trolling (e.g. students that label randomly other peers as friends/enemies). We hypothesize that isolated relationships continue to emerge until directed paths of length $2$ dominate the dynamics of network formation. As discussed in previous sections, paths of length $2$ are equivalent to intermediate students who can get two of their contacts in touch with each other. This mediation, quantified by the triadic influence, is an extremely good predictor of relationships, with accuracies as high as 86\%. Interestingly, when we focus on relationships that are not isolated (there are directed paths of length 2 connecting both students), prosociality is still a good predictor of them. This suggests that some of these relationships might have originated as isolated relationships, and that prosociality is still important even when the relationship is not isolated. Complementary to this, we have observed that the accuracy achieved by the triadic influence does not improve if we also provide personal information about the students. This implies that the triadic influence somehow subsumes the information on the students' characteristics, rendering it irrelevant to predict relationships. It is still an open question whether information obtained from more elaborated personality tests could improve on the predictions achieved by the triadic influence alone.

On the other hand, we have used state-of-the-art algorithms to create an embedding for each student that contains information about their surrounding, considering only the undirected and unweighted network. We have shown that this structural information can be used to predict the type of relationship between two students. The embedding of each node is created using a random walk exploration of its surrounding, the depth of which is a parameter that we can vary (see Methods). Depending on the typical length of the exploring random walks this method can gather different structural information. The maximum length of the random walks used in this study is $L=4$ (see the SI Fig.~S6). Therefore, the Node2Vec algorithm is exploring the local structure of each student. This aligns with the results achieved using the triadic influence, suggesting that the closest contacts in the network---the local environment---are the ones that influence the creation/transformation of relationships the most. Still, it is remarkable that the triadic influence---a one-dimensional metric---is able to achieve a higher accuracy than the prediction based on the embeddings. This suggests that the triadic influence gathers more information than the metrics included in the embedding of the nodes---although note also that we are using different information in both cases because the embeddings are created irrespective of the weights and directions of the relationships.

Interestingly, Ref.~\cite{boardman2012} suggests that individuals with similar genotypes may not actively select into friendships. Instead, they may be placed into these contexts by institutional mechanisms outside of their control. Our conclusions could be interpreted in a similar way; the triadic influence may act as a social force that encourage students that are compatible (incompatible) to have positive (negative) relationships, akin to the popular knowledge ``to be judged by the company you keep''. In this case, prosociality would be still a good predictor of the relationship even though it was the social context---the triadic influence in our case---which promoted the relationship. This raises an important point that we want to stress: predictability does not imply causality. Another situation that highlights the difficulty of disentangling cause and effect is that at the time we collected the data many relationships that nucleated in isolation due to prosociality alone were now surrounded by multiple directed paths of length $2$, and we have shown that the triadic influence is a very good predictor of the label of these relationships, even if their existence predated the paths entering the computation of the triadic influence. Therefore, while our results suggest a nucleation mechanism based on individual traits followed by a growth and evolution of the network dominated by the triadic influence, they do not prove that this is indeed the case. In order to assess to what extent this idea describes what is actually happening in real networks, a possibility would be to use the probabilities that we have learned through our link prediction techniques to simulate growing/evolving networks, and then compare these simulations with real data. In particular, it will be extremely interesting to collect data for the same network at different times to test the plausibility of different mechanisms of network evolution based on the probabilities learned here. If our proposal remains a good candidate to explain how networks form and evolve, then specific questions of interest arise, such as when the paths of length $2$ begin to dominate over the primitive relationships existing in a network or how a local change in the sign of a relationship can lead to a cascade of changes with global effects on the social network.

Finally, it is worth mentioning that our results come from data from a large number of surveys but from a very specific population, namely, teenagers in secondary schools in Spain. Thus, the generality of our results should be validated by gathering similar data from other collectives and performing similar analyses.

\section*{Methods}

\subsection*{Data collection}

Surveys were conducted in 13 Spanish high schools (mandatory education, 11 to 15 years of age). The surveys were delivered through a computer interface and included direct questions about their relationships, as well as some others aimed at identifying personal attributes. To elicit relationships, students could choose from a list containing all the other students in their same year within their own schools. The number of classes participating in the study in each school depended on the availability of time and the decisions of the school direction. The data corresponding to one of the schools, also included in this work, was presented in full detail in Ref.~\cite{escribano2021}. For each student, we collected:
\begin{itemize}
    \item \textbf{General data:} School ID, course, class, and a student ID assigned by the software for the purpose of this study. 
    \item \textbf{List of relationships:} All the relationships declared by the student (very good, good, bad and very bad) were collected with the student IDs of the nomenees and the corresponding labels ($+2$, $+1$, $-1$, $-2$). 
    \item \textbf{Individual traits:} 
    \begin{itemize}
        \item Gender, which included 1789 males, 1720 females, and 4 non-binary people. 
        \item Cognitive reflection test (CRT), computed using the answer to 3 questions about logic \cite{branas-garza2019,branas-garza2022}, and yielding values $0$, $1$, $2$ and $3$.
        \item Prosociality, evaluated through the answer to the three following questions about sharing ($q_i$ ranks the level of selfishness of each answer):
    \begin{itemize}
        \item What do you prefer? A) 10€ for you and 10€ for your partner ($q_1=0$) B) 10€ for you and 0€ for your partner ($q_1=1$).
        \item What do you prefer? A) 10€ for you and 10€ for your partner. ($q_2=1$)  B) 10€ for you and 20€ for your partner ($q_2=0$).
        \item What do you prefer? A) 10€ for you and 10€ for your partner ($q_3=0$) B) 20€ for you and 0€ for your partner ($q_3=1$).
    \end{itemize}
    The selfishness score is $s=q_1+q_2+q_3$, and the prosociality index is obtained as $p=1-(s/3)$.
    \end{itemize}
\end{itemize}


\subsection*{Predicting relationships using local information}

Our social networks are directed graphs representing the relationships between all the students within each of the high schools of our study. We kept only the students that answered all the tests about their individual features (described above), a total of 3395 students and 60566 relationships. Relationships are gathered in the weighted adjacency matrix $W$, with elements $w_{ij}\in\{ -2, -1, 0, 1, 2\}$ corresponding to the value of the relationship that student $i$ declares to have with student $j$ ($w_{ij}=0$ if there is no declared relationship). Note that $w_{ii}=0$ and that $W$ is not symmetric (relations are not necesarily reciprocal). Additionally, the individual traits described above (self-declared gender, CRT, and prosociality) are stored in the nodes $n_i$ of the graph. A key quantity used in this work is the triadic influence $I_{ij}\equiv(W^2)_{ij}=\sum_kw_{ik}w_{kj}$. It quantifies the aggregated contribution of the directed paths of length $2$ that go from $i$ to $j$.  Note that triadic influence considers only \textit{directed} paths from $i$ to $j$, and that $I_{ij} \ne I_{ji}$ in general.

In order to use a neural network to predict the declared relationships between students, we would like to avoid having highly unbalanced classes, and therefore we define a task with only two classes: friends (we consider here only $+2$ relationships) or enemies (we merge here relationships $-2$ and $-1$). We have also considered a more unbalanced case, with the friend class corresponding to relationships with labels +1 and +2 and the results were qualitatively analogous. In any case, when we compute the triadic influence $I_{ij}$, we keep all the labels in the network $\{-2,-1,1,2\}$ (see Fig.~\ref{fig_diagram} for an example). In this section, we use a deep neural network with one hidden layer, ReLu activation (see e.g.~Ref.~\cite{goodfellow2016}), and $100$ hidden units. The input dimension depends on the data we want to use to predict the relationship. Our neural network is a nonlinear function of the inputs and the internal parameters (numbers that change their value during training), which outputs a vector of dimension two. Let us call these outputs $f(\mathcal{I},\mathcal{W})_i$, where $\mathcal{I}$ stands for the inputs corresponding to one specific relation (triadic influence, gender of both students \dots), $\mathcal{W}$ are the internal parameters of the network and $i=0,1$ indicates one of the two classes in our dataset (friends/enemies). Then these outputs are put into a SoftMax function (see e.g. Ref. \cite{goodfellow2016}) such that
\begin{equation*}
    q(\mathcal{I},\mathcal{W})_i \equiv
    \frac{e^{f(\mathcal{I},\mathcal{W})_i}}{e^{f(\mathcal{I},\mathcal{W})_0}+e^{f(\mathcal{I},\mathcal{W})_1}},
\end{equation*}
where $q(\mathcal{I},\mathcal{W})_i$ can be interpreted as the probability that a specific sample, characterized by inputs $\mathcal{I}$, belongs to class $i=0,1$. Training the neural network amounts to minimizing a loss function such that $q(\mathcal{I},\mathcal{W})_i$ resembles the \emph{actual} probability distribution $p(\mathcal{I})_i=\delta_{i,\ell(\mathcal{I})}$ for each sample---$\ell(\mathcal{I})$ being the label of that input data and $\delta_{i,j}=1$ if $i=j$ and $0$ otherwise. We use the cross-entropy loss function
\begin{equation*}
\begin{split}
    \mathcal{L} &= -\sum_{k,i} p(\mathcal{I}_k)_i \log( q(\mathcal{I}_k,\mathcal{W})_i)=
    -\sum_k \log( q(\mathcal{I}_k,\mathcal{W})_{\ell(\mathcal{I}_k)}),
\end{split}
\end{equation*}
where the index $k$ runs over all samples in the dataset. Note that if $q(\mathcal{I}_k,\mathcal{W})_{\ell(I_k)}=1$ for all $k$, the network would predict with $100$\% certainty the correct label for all samples. In this situation $\mathcal{L}=0$, indicating that for the set of parameters $\mathcal{W}$ the function $\mathcal{L}$ reaches an absolute minimum. Hence, training the neural network amounts to minimizing $\mathcal{L}$ with respect to the parameters $\mathcal{W}$. We have used stochastic gradient descent with an initial learning rate of $0.1$ and a decaying factor of $0.99$. We use a minibatch of size $20$ and oversample the class with the smallest number of samples so that each minibatch has the same number of samples from each class. Unless otherwise stated, we minimize for $200$ steps and compute the accuracy in the final step. In the case of the prediction of isolated relationships (two bottom bars of Fig.~\ref{fig_acc}), the dataset is greatly reduced. To ensure that our results are robust we use a 10-fold cross-validation approach and report the mean value and an error bar representing the standard deviation from the mean. In this case, we train for $1000$ minimization steps using a dynamical loss function with oscillations of amplitude $10$ and a period of $5$ minimization steps. A dynamical loss function weights the contribution of each class to the loss function with proportionality factors that oscillate during minimization. This process changes the topography of the loss function landscape \cite{ruiz2019tuning}, and helps the model find deeper and wider minima of the loss function (see Ref.~\cite{ruiz2021tilting} for further details).

\subsection*{Predicting relationships using global information}\label{global_information}

The steps followed in the process of creating the embeddings and predicting the class of a relationship are: 
\begin{itemize}
    \item Passing the graph as an object to Node2Vec \cite{grover2016node2vec} yields a $128$-dimensional vector for each node (an embedding). Node2Vec is defined by the two hyperparameters $(p,q)$, which describe the space explored by the random walks. We use $(p=1,q=4)$ after doing a hiperparameter optimization. The characterization of the typical random walk in this process can be found in the Supporting Information.
    \item We merge the embeddings for each pair of nodes to create the embedding of each edge (relationship), $\mathbf{e}$.
    \item The structural representation for each edge, $\mathbf{e}$, is the input that we use to predict the label, friends/enemies of the relationships in the training dataset. We oversample the training data (test data are left untouched) using the SMOTE technique \cite{chawla2002smote}. This method produces new examples by interpolating close existing points in the 128-dimensional space.
    \item We apply two different machine learning procedures: a random forest, and an artificial neural network. 
\end{itemize}

The artificial neural network was implemented in the standard library Tensorflow \cite{tensorflow2015-whitepaper} with four hidden layers (sizes $128$, $64$, $32$, and $8$) and the ReLu activation function. The final output included a sigmoid function.

\begin{acknowledgments}
This work has been partly supported by grant PGC2018-098186-B-I00 (BASIC), funded by MCIN/AEI/10.13039/501100011033 and by ``ERDF A way of making Europe''. M.R.-G. acknowledges support from the CONEX-Plus program funded by Universidad Carlos III de Madrid and the European Union's Horizon 2020 research and innovation program under the Marie Sk{\l}odowska-Curie grant agreement No.~801538. P.B.-G.~acknowledges support from Junta de Andaluc\'{\i}a (PY18-FR-007) and Agencia Andaluza de Cooperaci\'on Internacional para el Desarrollo (AACID-0I008/2020) and MCIN (PID2021-126892NB-I00).
\end{acknowledgments}

\appendix
\section{Statistical analysis of the data}

Our dataset compromises $13$ schools with $3395$ students and $60566$ declared relationships, where we have already removed the students that did not answer all the questions about personality (around 3\% of the students) and the relationships that included them. From the $13$ schools considered, $3$ of them are in the Region of Madrid and the rest are in Andaluc\'ia. In this Supplementary Materials we will name the schools in Andaluc\'ia as t11\_1,  \dots, t11\_10, whereas the schools in Madrid will be t1, t2 and t6. 

\begin{figure}[h!]
    \centering
    \includegraphics[width=0.7\linewidth]{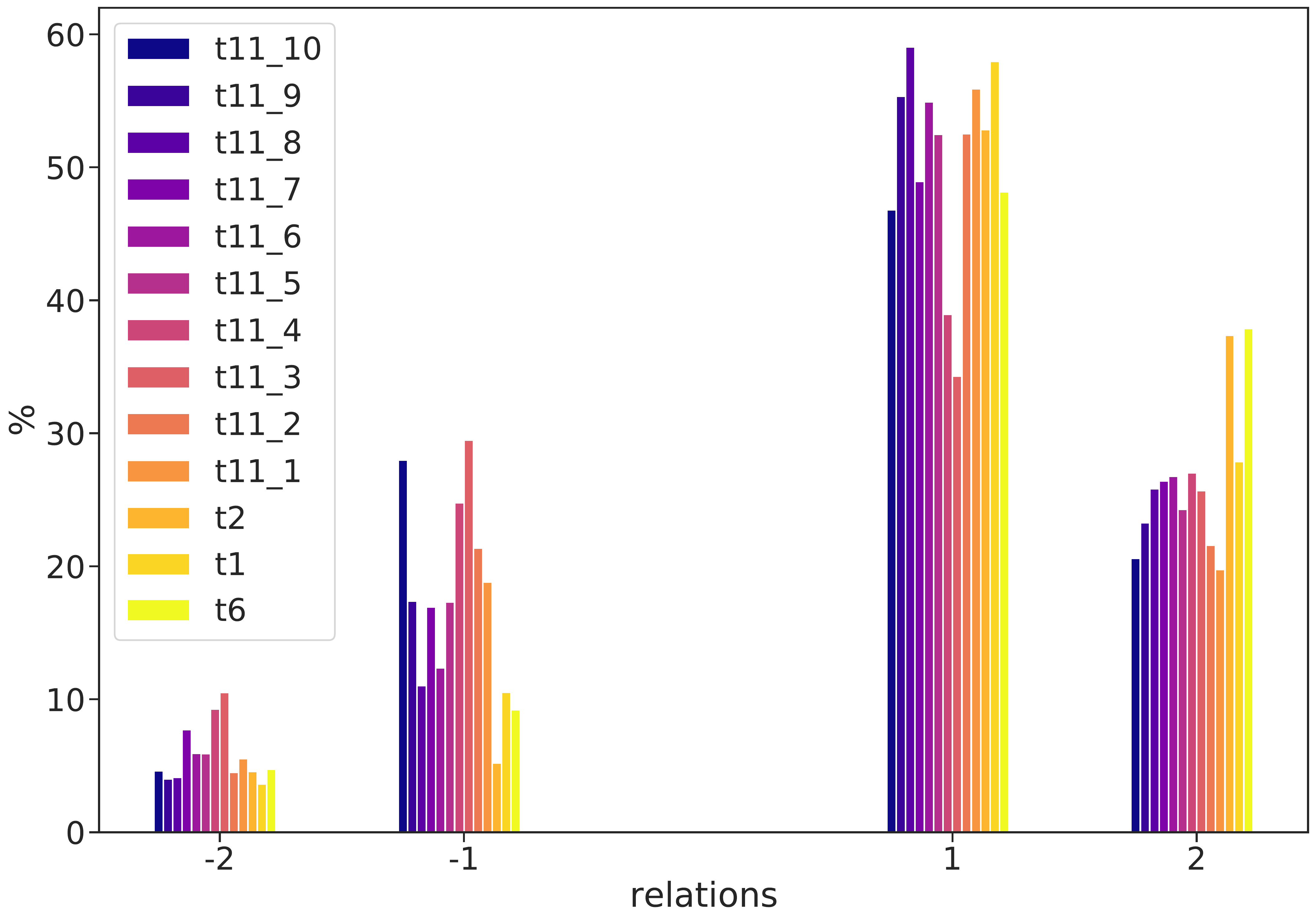}
    \caption{Proportion of the different types of relationship at each school. Students declare relationships with values $-2, -1, 1, 2$, from nemesis to best friends. These relationships are directed, from student $i \rightarrow j$ with value $W_{ij}$. We have studied $13$ schools with $3395$ students in total, and $60566$ declared relationships. Most common relationships are $+1$ followed by $+2$, $-1$ and $-2$. We observe similar percentages across schools. }
    \label{fig_relationships}
\end{figure}

Students can declare to have a very good ($+2$), good ($+1$), bad ($-1$) or very bad ($-2$) relationship with any other student in their school. The distribution of the four types of relationships changes among schools although some features are common, see Fig. \ref{fig_relationships}. For all schools, the most numerous relationships are good ($+1$) relationships, accounting between $40$ and $50$\% in most schools. On the contrary, the least abundant relationships are the very bad ones ($-2$). In addition to this, we see some differences between the schools in Madrid and Andaluc\'ia but we leave a detailed study of these features for a future work.

\begin{figure}
    \centering
    \includegraphics[width=0.7\linewidth]{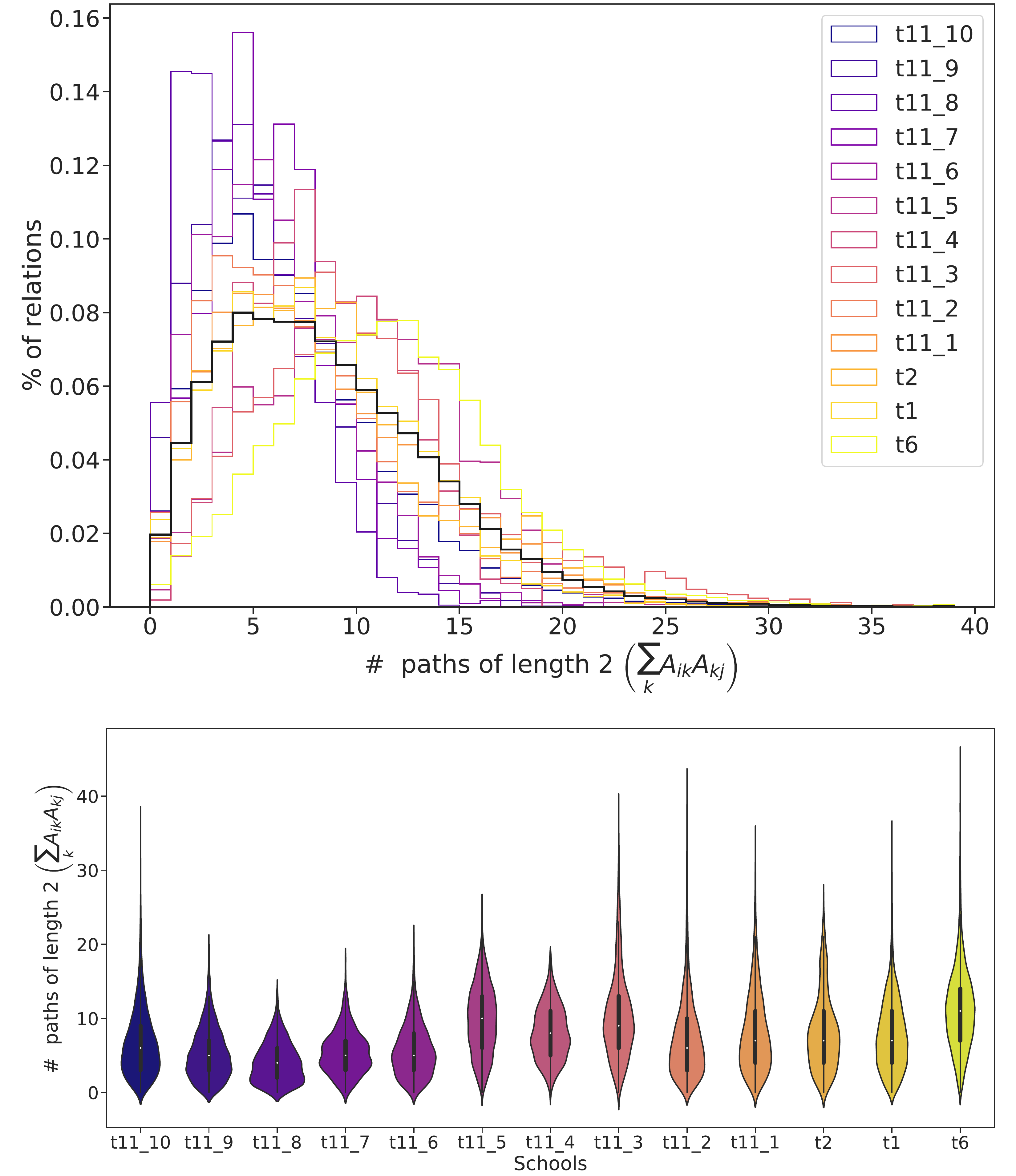}
    \caption{Percentage of relationships according to the number of paths of length 2 connecting the same nodes. For each relationship from $i$ to $j$ we compute the number of paths of length 2 that go from $i$ to $j$, after this we compute the number of relations with a specific number of length-2 paths. Upper panel displays a color line for each school (see legend) and a thicker black line for the distribution corresponding to the complete dataset (clustering all the schools together). Bottom panel show a violin plot for each school to ease visualization.}
    \label{fig_paths}
\end{figure}

According to the main text, the most important information to predict a relation between students $i$ and $j$ is the triadic influence. This quantity accounts for the influence that third people have on other relationships, and uses the directed paths of length $2$ that connect $i$ to $j$. For this reason, it is interesting to know how many relationships have a particular number of paths of length $2$ connecting the starting node ($i$) and the final node ($j$). Figure \ref{fig_paths} encodes this information. We have plotted each school separately (color lines) and all the schools together (black line). From the plot, we can see that $2$\% of all relationships do not have any path of length $2$ connecting the starting and final node. The violin plot helps to visualize each school separately. Although there is variation between schools, the peak of the joint distribution (black line) is around $5$ or $6$ paths of length $2$.

\begin{figure}
    \centering
    \includegraphics[width=0.7\linewidth]{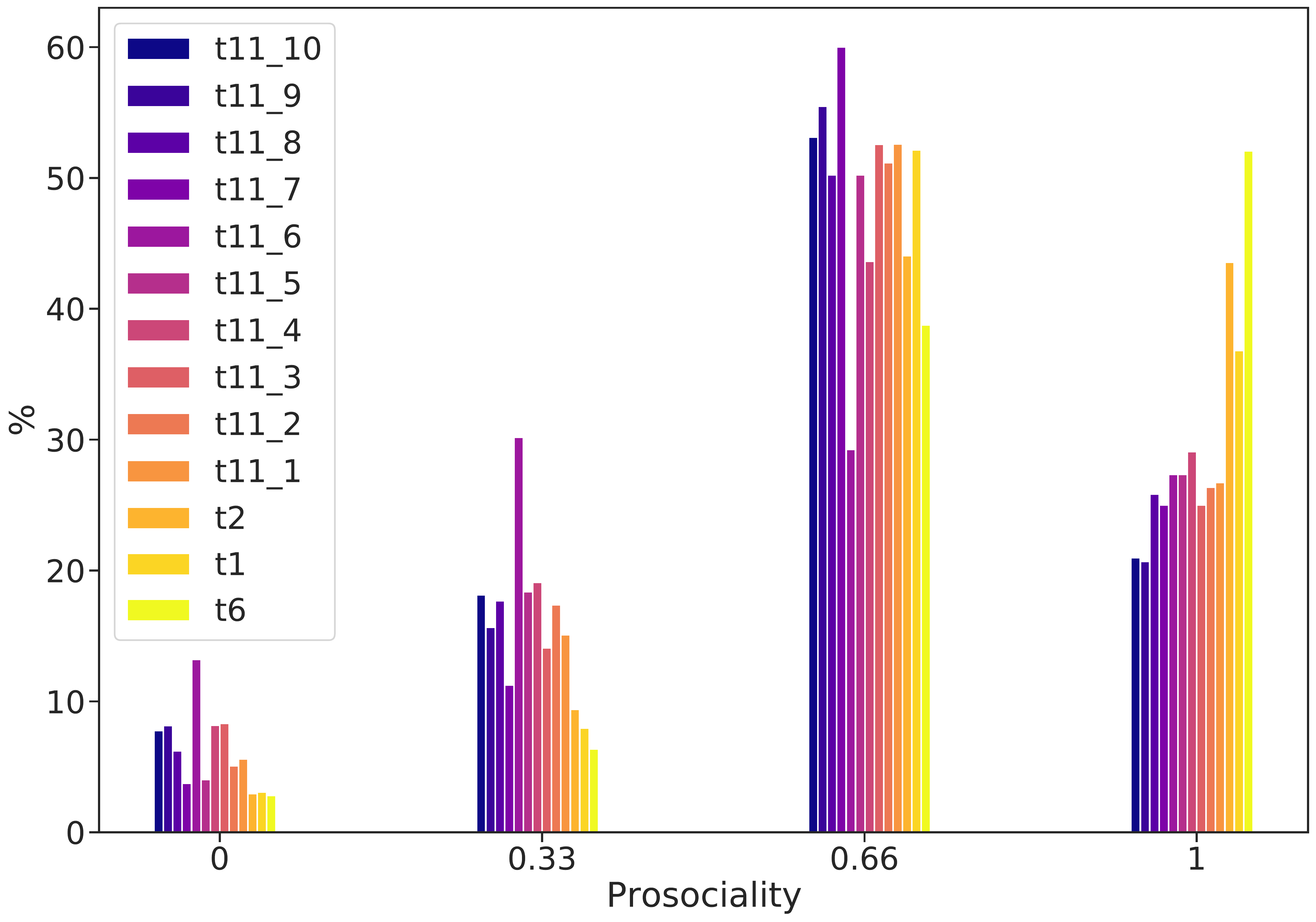}
    \caption{Distribution of prosocial behavior across schools. Prosocial behavior takes values $0, 0.33, 0.67, 1$ and bars correspond to the percentage of students in each school that display a certain level of prosociality. Distributions are similar across schools although t1, t2 and t6 (schools in Madrid) seem to have a larger proportion of very prosocial students ($1$) when compared to the other schools from Andaluc\'ia. }
    \label{fig_prosocial}
\end{figure}

Prosociality is computed for each student. Students answer 3 questions that lead to a scalar that takes values $0, 0.33, 0.66 \text{ or } 1$ (see Methods). The proportion of students per prosociality value is similar between schools, see Fig. \ref{fig_prosocial}, and resembles the distribution of relationships (Fig. \ref{fig_relationships}). The largest group of students in most schools is students with prosociality $0.66$, while the smallest group corresponds to antisocial students ($0$). Again, there is an apparent difference between the schools in Andaluc\'ia and Madrid. The latter seem to have a larger proportion of very social students $1$ compared to the schools in Andaluc\'ia. This seems to be in good agreement with the distribution of relationships where schools in Madrid showed more very good relationships ($+2$). We leave a detailed study of this phenomenon for future work.

\begin{figure}
    \centering
    \includegraphics[width=0.7\linewidth]{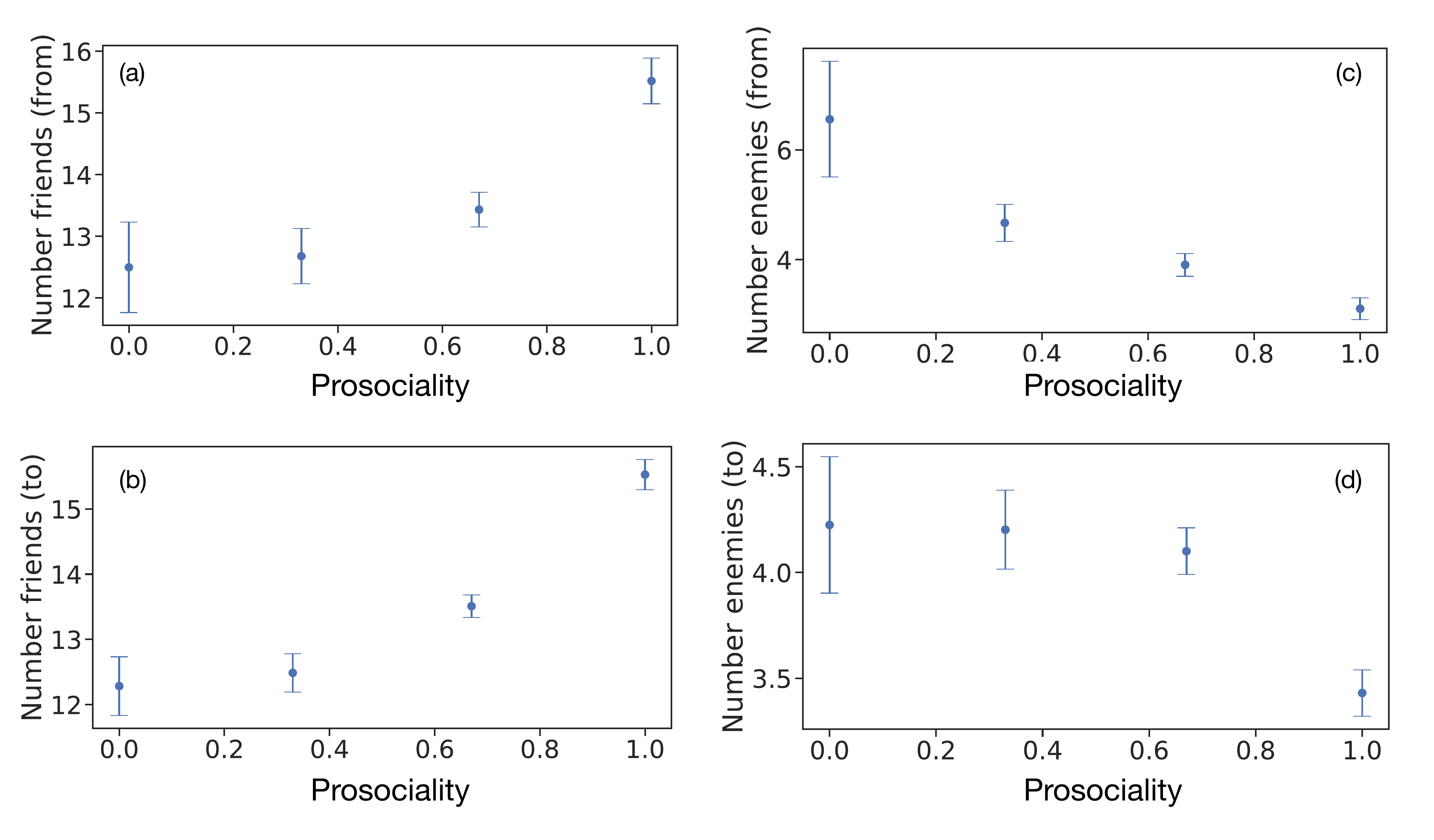}
    \caption{Statistics for the number of friends and enemies depending on the prosociality of the students. We merge here $-2$ and $-1$ relationships into an enemy category and $+2$ and $+1$ relationships as friends. Panel (a) shows the average number of people nominated as friends by a student of a specific prosociality, whereas panel (b) shows the average number of people that nominate someone of a specific prosociality as a friend. Panels (c) and (d) are analogous to (a) and (b) but for enemies. Students with high (low) prosociality tend to nominate and be nominated as friends (enemies) more frequentely. Error bars correspond to the error of the mean.}
    \label{fig_statistics}
\end{figure}

Figure \ref{fig_statistics} shows the average number of friends/enemies that are nominated by (or that nominate to) students of different prosociality values. Clear general trends indicate that students of high prosociality nominate and are nominated as friends more often, whereas students with low prosociality belong to negative relationships in a larger proportion than highly prosocial students.  These results are in good agreement with the probabilities learnt by the neural network in the main text, where students with low prosociality had a larger probability of being enemies whereas highly prosocial students have a larger probability of being friends.

\section{Predicting using only the structure of the network: additional information on the creation of the embedding and alternative results using a Random Forest}
\begin{figure}
\centering
\includegraphics[width=0.5\linewidth]{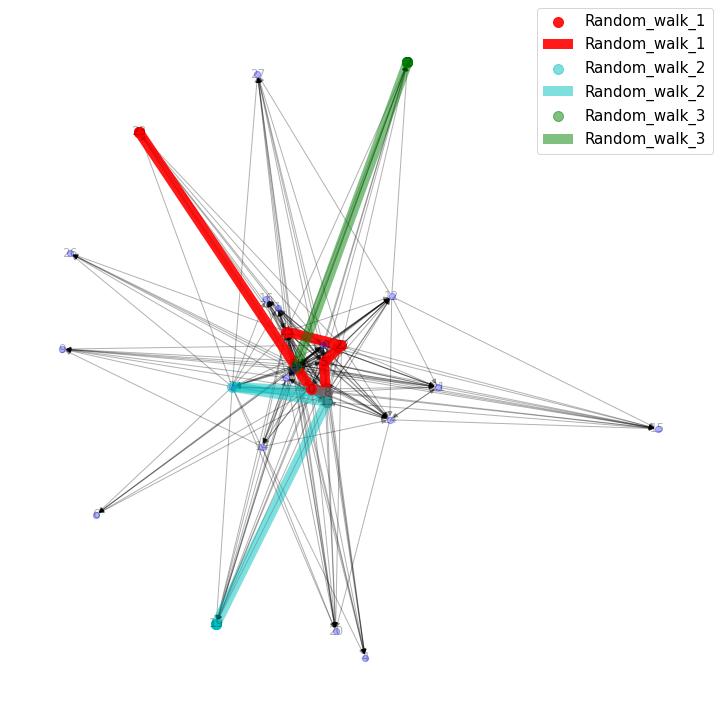}
\caption{\label{fig:RW_examples} Network representation of the students in one of the classes in the dataset, together with three examples of random walks used by \textit{Node2Vec} to create the embedding of one node. The random walks are controlled by the hyperparameters (in our case $p=1$ and $q=4$), they measure the probability of exploring or staying in a certain node along the path. From the figure we can see that Node2Vec is exploring the local environment of the node. }
\end{figure}

\begin{figure}
\centering
\includegraphics[width=0.7\linewidth]{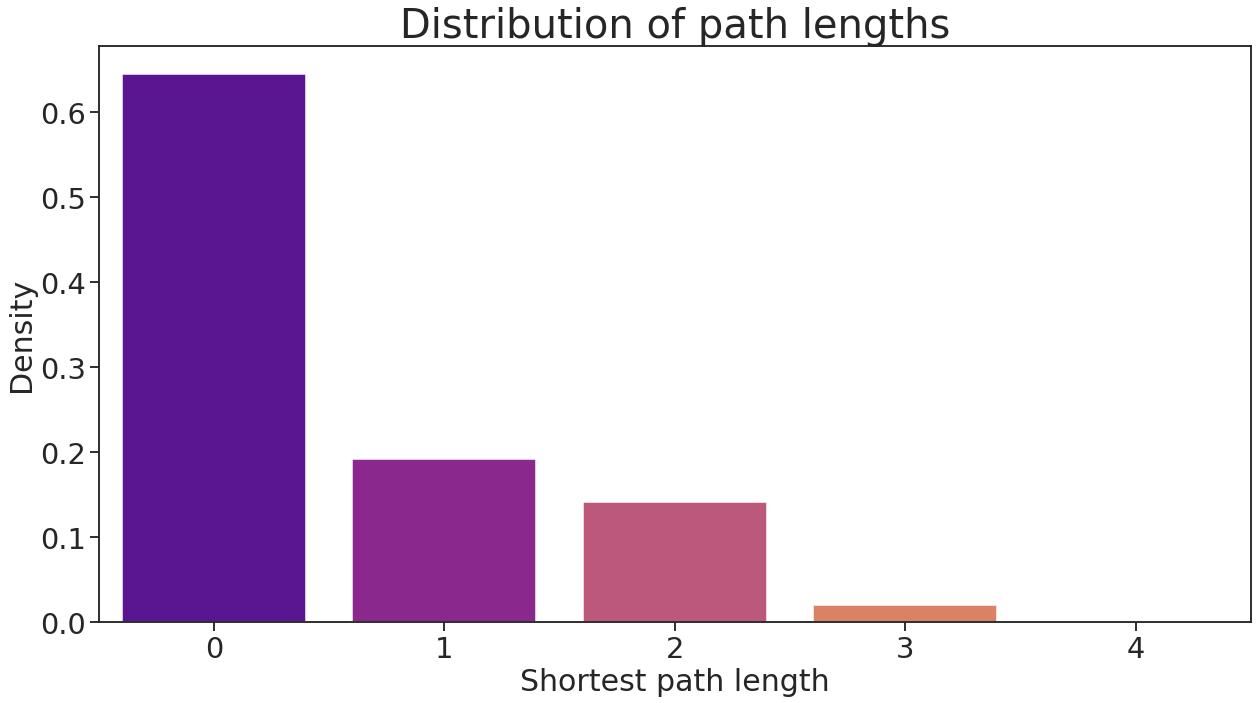}
\caption{\label{fig:RW_dist} Distribution of path lengths in the random walks used by Node2Vec. This length is the shortest path between the origin of the random walk and the farthest node in that walk. A walk of length $0$ means staying at the same node. We can therefore understand the locality of the algorithm. The shortest path length has an average and standard deviation of $0.80\pm 1.07$, which shows again that embeddings are mainly built with local structure.}
\end{figure}

We include here some additional information to clarify how the Node2vec algorithm works. Node2Vec builds the embedding for each node using random walks. The configuration of hyperparameters that we have chosen ($p=1$ and $q=4$) leads to local exploration. To create the embedding of one node we use 420 random walks, each of them composed of 30 movement attempts. In order to get some intuition, we show in Fig.\ref{fig:RW_examples} the path followed by three random walks starting from different nodes, we show RWs that does not stay fixed in the initial node, although that is the most probable situation. The local exploration of the surrounding of the node by the random walks is confirmed by Fig.\ref{fig:RW_dist}, which shows the distribution of lengths for the random walks with the hyperparameters used in our work (see Methods).

\begin{figure}
\centering
\includegraphics[width=0.7\linewidth]{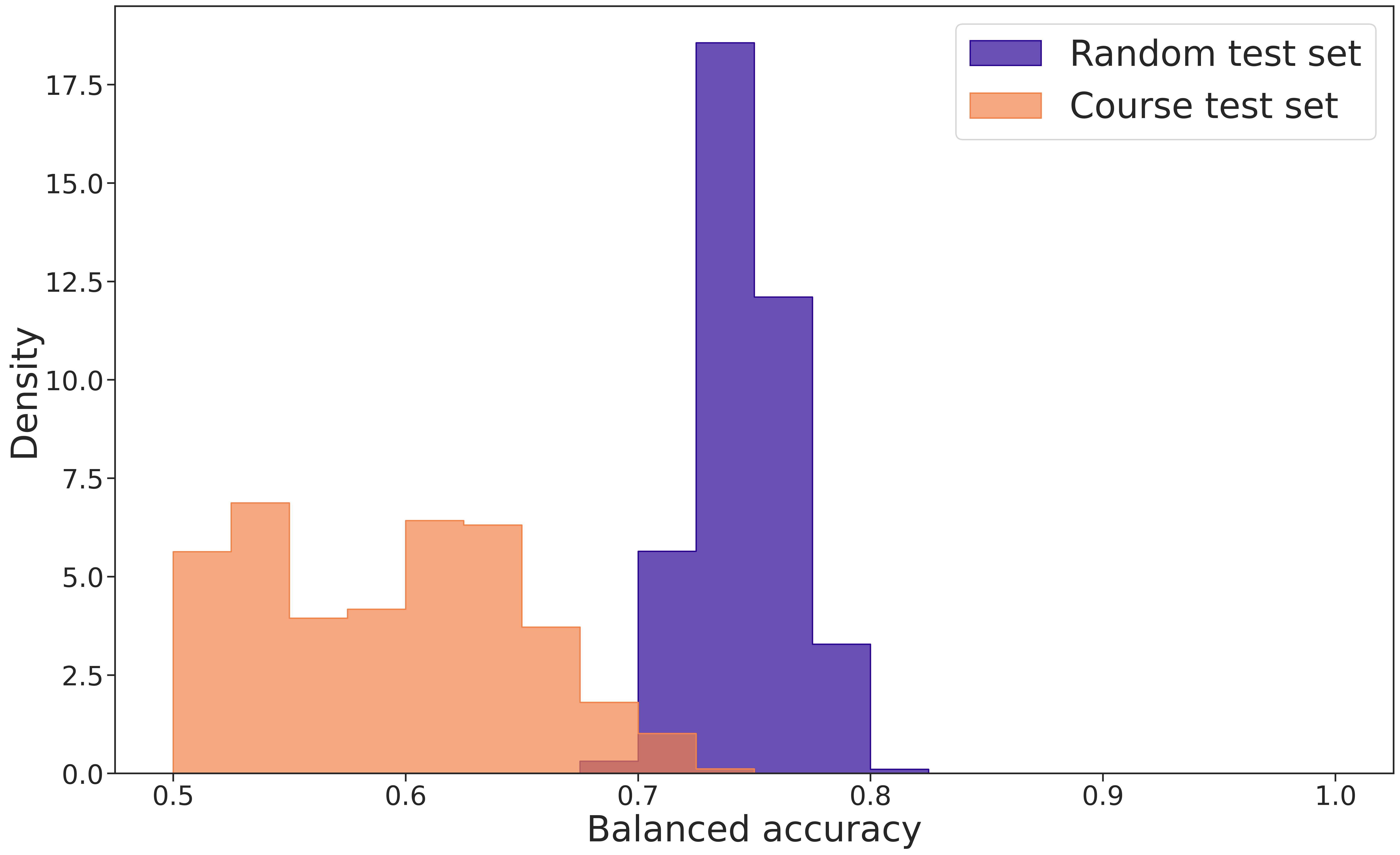}
\caption{\label{fig:AUC_RF} Predictions achieved with a Random Forest. Distribution of balanced accuracy for the 13 high schools. Each histogram is composed of a sample of $N=390$ points, that are different simulations for the same treatment, and then normalized such that the area of the histogram sums 1. The purple/dark histogram represents \emph{treatment I} whereas the orange histogram represents \emph{treatment II}, that is, using an specific age level from a high school as the test set. }
\end{figure}

Finally, to complement the results shown in the main text, we also include here the results for a second method using the embeddings as input. We have used a Random Forest (scikit-learn \cite{scikit-learn}). A Random Forest is an algorithm that divides the data into random ensembles of predictors and data. From these randomly chosen pieces of data, the algorithm builds and trains decision trees. The final decision is then made on the most popular answer within this population of decision trees. We used a maximum depth of 7 levels for our decision trees. Fig. \ref{fig:AUC_RF} depicts qualitatively analogous results to the ones achieved with the neural network and shown in the main text.

\clearpage
\bibliography{aapmsamp}

\end{document}